\documentclass[preprint,11pt]{elsarticle}
\usepackage{amsmath,amsthm,amsfonts,amssymb}
\usepackage[mathcal]{eucal}
\usepackage{mathrsfs}
\usepackage{dcolumn}
\usepackage{latexsym}
\usepackage{bm}
\usepackage[all]{xy}
\usepackage[T1]{fontenc}
\journal{Physics Letters B}

\begin{document}
\begin{frontmatter}
\title{Vector and fermion fields on a bouncing brane with a decreasing warp factor in a string-like defect}

\author[ufc,ifce]{L. J. S. Sousa}
\ead{luisjose@fisica.ufc.br}

\author[ifpb]{C. A. S. Silva}
\ead{calex@fisica.ufc.br}

\author[ufc]{D. M. Dantas}
\ead{davi@fisica.ufc.br}

\author[ufc]{C. A. S. Almeida}
\ead{carlos@fisica.ufc.br}

\address[ufc]{Departamento de F\'{\i}sica - Universidade Federal do Cear\'{a} - UFC \\  C.P. 6030, 60455-760 Fortaleza - Cear\'{a} - Brazil
}

\address[ifce]{Instituto Federal de Educa\c{c}\~{a}o Ci\^{e}ncia e Tecnologia do Cear\'{a} (IFCE) - Campus de Canind\'{e} \\
62700-000 Canind\'{e} - Cear\'{a} - Brazil}

\address[ifpb]{Instituto Federal de Educa\c{c}\~{a}o Ci\^{e}ncia e Tecnologia da Para\'{i}ba (IFPB) - Campus Campina Grande \\
Rua Tranquilino Coelho Lemos, 671, Jardim Dinam\'{e}rica,\\ Campina Grande - Para\'{\i}ba - Brazil}

\begin{abstract}
In a recent work, a model has been proposed where a brane is made of a scalar field with bounce-type configurations and embedded in a bulk with a
string-like metric. This model produces an AdS scenario where the components of the energy momentum
tensor are finite and have its positivity ensured by a suitable choice of the bounce configurations.
In the present work, we study the issue of gauge and fermion field localization in this scenario. In contrast with the five dimensional case here the gauge field is localized without the dilaton contribution. Nevertheless, it is remarkable that the localization of the fermion field depends on the introduction of a minimal coupling with the angular component of the gauge field, which differs clearly from five dimensional scenarios. Furthermore, we perform a qualitative analysis of the fermionic massive modes and conclude that only left handed fermions could be localized in the brane.

\end{abstract}

%\maketitle

%\keywords{Field theories in higher dimensions, Standing waves, String-like defect}
\end{frontmatter}

\section{Introduction}

Thick branes have been proposed as a smooth generalization of the Randall-Sundrum scenario
\cite{Kehagias2001,Rubakov1983,Dzhunushaliev2010,Akama1983}. In this model,
five-dimensional gravity is coupled to scalar
fields. Thick brane models consist in a more realistic scenario than the Randal-Sundrum
one, since no singularities appear due to the form of the scalar potential functions.

As a matter of fact, thick brane models have been comprehensively used in the task of localization of physical fields on the brane.
The importance of this subject stays in the fact that the introduction of extra dimensions affects both gravitational
interactions and particle physics phenomenology, and leads to modifications in the standard
cosmology.  If the extra dimensions indeed exist, it will inevitably change our ideas about the
universe. The quest of field localization can guide us to which kind
of brane structure is more acceptable phenomenologically \cite{W.T.Cruz2009}.

In this context, gravitons and fermions, as well as gauge fields can be localized on the brane in thick brane models. Gauge fields, in particular,
are localized only with the help of the dilaton
field. The Kalb-Ramond field localization in this scenario was also studied by \cite{Cruz2009}. There the use of the dilaton was again necessary in
order to localize the Kalb-Ramond field on the brane.

On the other stand point, scenarios have been proposed where thick brane solutions are extended to spacetimes with dimension more than
five \cite{Dzhunushaliev2010}. Among these works, we have some where branes are embedded in a bulk with a string-like metric.
The mainly motivation to
study branes in the presence of a string-like bulk comes from the fact
that most of the Standard Model fields are localized on a string-like defect. For example, spin-$0$, spin-$1$, spin-$2$, spin-$1/2$ and spin-$3/2$ fields are all localized on a string-like structure. Particularly, the bosonic fields are localized with exponentially decreasing warp factor, and the fermionic fields are localized on the defect with increasing warp factor \cite{Oda2000}. Even more interesting is the fact that spin-$1$ vector \cite{Oda2000}, as well as the Kalb-Ramond field \cite{W.T.Cruz2009}, which are not localized on a domain wall in Randal-Sundrum model, can be localized in the string-like defect.

However, most of the thick brane models in six dimensional scenarios, proposed so far, have been suffering from some drawbacks. The first difficult is
related with the introduction of scalar fields as a matter-energy source in the equations. In this case it is very difficult to find analytical
solution to the scalar field and to the warp-factor as well. Koley and Kar \cite{Koley2007} have suggested a model where analytical solutions can be found in a six dimensional scenario, however they run into a second difficult. This difficult is related with the positivity of the components of the
energy-momentum tensor and has been found by other authors also \cite{Dzhunushaliev2010, Koley2007, Dzhunushaliev2008}. Finally, field localization are not possible for any field in the thick branes defined in Ref. \cite{Dzhunushaliev2008}, at least for physically acceptable solutions.

On the other hand, in a recent work \cite{Sousa:2012jw}, a model was proposed where a brane is made of a scalar field with bounce-type configurations and embedded in a bulk with a
string-like metric. This model produces a sound AdS scenario where none of the important physical quantities is infinite. Among these quantities
are the components of the energy momentum
tensor, which have its positivity ensured by a suitable choice of the bounce configurations.
Another advantage of this model is that the warp factor can be obtained  analytically from the equations of motion for the
scalar field, obtaining as a result a thick brane configuration, in a six dimensional context. It has been shown that scalar field localization is
suitable in the scenario proposed in Ref. \cite{Sousa:2012jw}, paving the way in the sense of localization of other fields.
Therefore, in the present work we will study the possibility of localization of vector and fermion
fields in these scenario, in order to test its applicability and robustness.

This paper is organized as follows. In section \ref{section2}, we address the model introduced in Ref. \cite{Sousa:2012jw}, where a bulk scalar
field with bounce-type configurations generates a brane which is embedded in a bulk with a string-like metric.
In section \ref{section3} we address the vector field localization, and in section \ref{section4} the fermion field localization is established.
On the other hand, section \ref{section4b} cope with qualitatively analysis of the fermionic massive modes. Section \ref{section5} is devoted to remarks and conclusions.

\section{The model} \label{section2}

The use of bulk scalar fields to generate branes was introduced by Goldberger and Wise
\cite{wd.goldberger-prl83, wd.goldberger-prd60}, and has been largely studied in the literature
\cite{o.dewolfe-prd62, rn.mohapatra-prd62, p.kanti-plb481, jm.cline-prd64, jm.cline-plb495, a.flachi-npb610}.
In the six-dimensional context,
Koley and Kar \cite{Koley2007} have built a scenario where the brane is made of scalar fields and
analytical ``thin brane'' solutions have been found out. Several progress have been obtained in the work
by Koley and Kar in the intend of construct
brane solutions in six dimensions, as well as, in the task of localize physical fields. However, some troubles with the energy conditions
(WEC, SEC, NEC) \cite{m.visser-aip} were found. In this model, the energy momentum tensor violates all the energy
conditions since its components are not positive defined.

Recently, an AdS type solution was found in a model which
assumes a six dimensional action
for a bulk scalar field in a  double well $V(\phi) = \frac{\lambda}{4}(\phi^{2} - v^{2})^{2}$ potential
minimally coupled to gravity in the presence
of a cosmological constant \cite{Sousa:2012jw}. In this scenario, which will underly the present article,
it is admitted that
the scalar field equation possess bounce-like statics
solutions depending only on the radial extra dimension, where the simplest is $\phi(r) = v \tanh(ar)$.

The model is described by the action
\begin{equation}
S=\frac{1}{2\kappa_{6}^{2}} \int d^{6}x \sqrt{-\;^{(6)}g}\Big[(R - 2\Lambda) + g^{AB}\nabla_{A} \phi \nabla_{B}\phi - V(\phi)\Big]\;, \label{action}
\end{equation}
where $\kappa _{6}$ is the $6$-dimensional gravitational constant, and $\Lambda$ is the bulk cosmological constant.

The fields live in a string-like scenario with the following metric	
\begin{eqnarray}
\lefteqn {ds^{2}=g_{MN}dx^{M}dx^{N}}\nonumber\\
& &=g_{\mu\nu}dx^{\mu}dx^{\nu}+\tilde{g}_{ab}dx^{a}dx^{b}\nonumber\\
& &=P\hat{g}_{\mu\nu}dx^{\mu}dx^{\nu}+dr^{2}+Q d\theta^2\nonumber\\
& &=e^{-A(r)}\hat{g}_{\mu\nu}dx^{\mu}dx^{\nu}+dr^{2}+e^{-B(r)}d\Omega_{(5)}^{2}, \label{sl-metric}
\end{eqnarray}
where $M,N,...$ denote the $6$-dimensional space-time indices, $\mu, \nu, ...,$ the $4$-dimensional brane ones, and $a, b, ...$ denote the $2$-extra spatial dimension ones. Also $d\Omega_{(5)}^{2}=R_{0}^{2}d\theta^{2}$, $P = e^{-A(r)}$ and $Q=R_{0}^{2} e^{-B(r)}$.

In the same way of the model introduced by Koley and Kar, the one introduced in \cite{Sousa:2012jw}
has the advantages to be analytical. However,
the introduction of the bounce-type configurations to the scalar field that generates the brane supports a way to solve the problems
with the energy conditions since the energy density may be positive or negative on the brane depending on the choice
of the bounce configurations.

Moreover, the finiteness of the relation between the four ($M_{p}$)
and six ($M_{6}$) dimensional reduced Plank scale \cite{Gherghetta2000} is ensured by the form of the warp factor
found out by the authors which is given by
\begin{equation}
A(r) = \beta \ln \cosh(ar) + \frac{\beta}{2} \tanh^{2}(ar) \label{warp-factor}\;.
\end{equation}

\noindent In the expression above
$\beta = \frac{1}{3} \kappa_{D} ^{2} \nu ^{2}$. Moreover, the warp factor found out
is equal to 1 at $r = 0$ which ensures that on the brane one has a 4D Minkowski space-time. Besides, as $r$ goes to
zero or infinity, the warp factor goes to $0$.

In the reference \cite{Sousa:2012jw}, the authors also pointed the interesting possibility of localization of  the standard model fields in this scenario.
In this way, the scalar field localization has been implemented paving the way for the localization of other fields.
An interesting result is that any non-gravitational trapping mechanism has not been necessary to localize scalar field
in this model, which can been seen as an advantage when compared with results of Dzhunushaliev and Folomeev \cite{Dzhunushaliev2008}.

In the present work, we will deal with vector and fermionic fields localization.
It is know that it is possible to localize chiral fermions in the ``5D version of this model'' \cite{Kehagias2001}.
However, to localize vector field in this set up, in five dimensions, we need to have a dilaton field present in the model forming a
``bounce-gravity-dilaton system" \cite{Kehagias2001}.
We expect, in this work, to localize
fields in this scenario that is more realistic than the RS model ones, without the necessity of the dilaton field.

\section{Localization of the vector field} \label{section3}
\label{vetor-local}

In this section we will address the issue of the vector field localization. As we will see, it is possible to localize the vector
field in this scenario if one has an exponentially decreasing warp factor, without the necessity of the dilaton field, as in the case of the the
scalar field discussed in Ref. \cite{Sousa:2012jw}.

In order to deal with the vector field localization, let us introduce the action
\begin{equation}
S_{m} = -\frac{1}{4} \int{d}^{D}x \sqrt{-g} g^{MN}g^{RS}F_{MR} F_{NS},
\end{equation}
where $F_{MN} = \partial_{M}A_{N} - \partial_{N}A_{M}$.

From the action above, one obtains the equation of motion
\begin{equation}
\frac{1}{\sqrt{-g}} \partial_{M}(\sqrt{-g} g^{MN}g^{RS}F_{NS}) = 0,
\end{equation}
which results in
\begin{equation}
\eta^{\mu\nu}g^{MN}\partial_{\mu}F_{\nu N} + e^{A + \frac{B}{2}} \partial_{r} \left( e^{- \frac{A + B}{2}}g^{MN}F_{rN} \right) + R_{0}^{-1}e ^{B - A}g^{MN} \partial_{\theta}F_{\theta N} = 0.
\end{equation}
This equation can be written in terms of equations for the vector field components as follows
\begin{equation} \label{vector-eq1}
\left( \eta^{\mu\nu} \partial_{\mu}\partial_{\nu} + e^{B/2}\partial_{r} e^{-(A + \frac{B}{2})}\partial_{r} + \frac{e^{B - A}}{R_{0} ^{2}} \partial _{\theta} ^{2} \right) A_{\lambda} - e^{B/2} \left(\partial_{r} e^{-(A + \frac{B}{2})} \partial_{r} \right) \partial_{\lambda}A_{r} = 0\;,
\end{equation}
\begin{equation} \label{vector-eq2}
\left(\eta^{\mu\nu}\partial_{\mu}\partial_{\nu} + \frac{e^{B - A}}{R_{0} ^{2}} \partial _{\theta} ^{2}\right) A_{r} = 0,
\end{equation}
and
\begin{equation} \label{vector-eq3}
\partial_{r} \left( e^{-2 A + \frac{B}{2}}\partial_{\theta}A_{r} \right) = 0.
\end{equation}
As has been done in \cite{Oda2000}, if we choose the gauge condition $A_{\theta} = 0$ and assume the decomposition
\begin{equation} \label{decomp-a-mi}
A_{\mu}(x^{M}) = a_{\mu}(x^{\mu}) \sum{\rho_{m} e^{il\theta}},
\end{equation}
and
\begin{equation} \label{decomp-a-r}
A_{r}(x^{M}) = a_{r}(x^{\mu}) \sum{\rho_{m} e^{il\theta}},
\end{equation}
we can show that there exist the s-wave (l = 0) constant solution $\rho_{m} = \rho_{0} = constant$ and $a_{r} = constant$. Note that we assume that $\partial_{\mu}a^{\mu} = \partial^{\mu}f_{\mu\nu} = 0$, where $f_{\mu\nu}$ is defined by $f_{\mu\nu} = \partial_{\mu}a_{\nu} - \partial_{\nu}a_{\mu}$.

\begin{figure}[htb] %
 \vspace{0pt}
       \begin{minipage}[t]{0.48 \linewidth}
                 \fbox{\includegraphics[width=\linewidth]{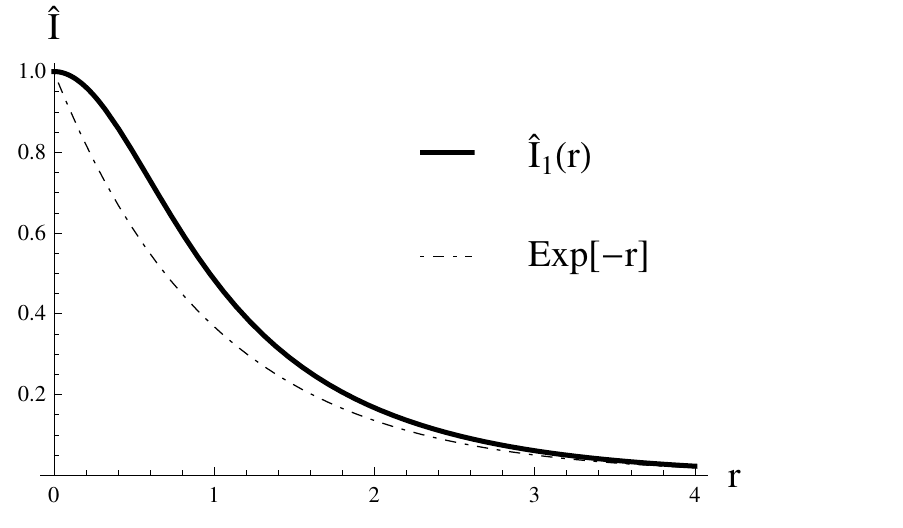}}\\
           \caption{Sketch of the integrand in eq.\eqref{iia1}. We can verify that this is a case of a smooth linear decreasing exponential. Here $\beta=a=1$.}
           \label{fig.ia1}
       \end{minipage}\hfill
       \begin{minipage}[t]{0.48 \linewidth}
                \fbox{\includegraphics[width=\linewidth]{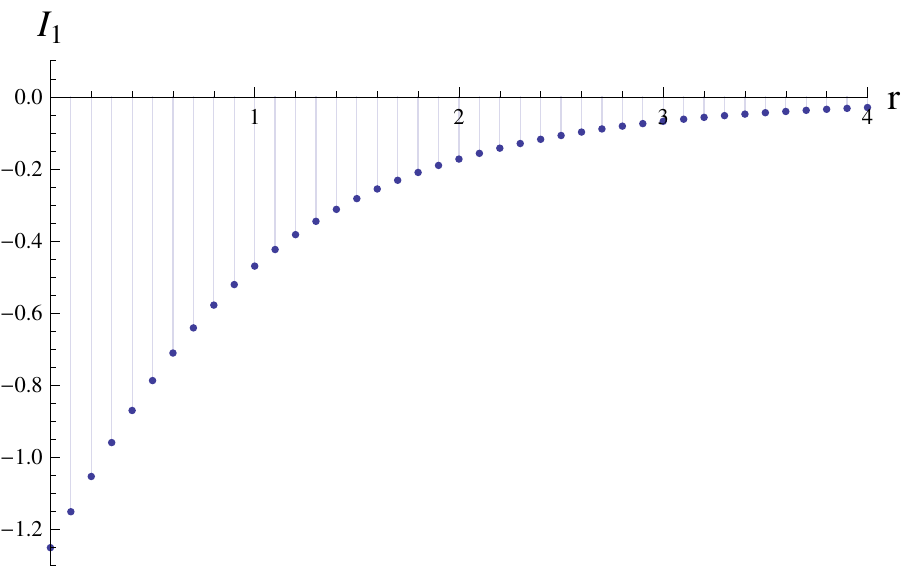}}\\
           \caption{Proof of convergence of $I_{1}$ (Eq.\eqref{iia1}). The function $I_{1}$ has a form similar to a linear decreasing exponential. After integration $I_{1}$ worth approximately $1.243$ for $\beta=a=1$.}
           \label{fig.i1}
       \end{minipage}
\end{figure}

By substituting the constant solution in the initial action, the resultant integral in the variable $r$ is
\begin{equation}\label{iia1}
I_{1} \propto \int_{0} ^{\infty}{dr e^{- \frac{B}{2}}}.
\end{equation}
In order to have zero mode localization for the vector field in this model, we need that $I_{1}$ to be finite. It is clear that in the case for $A=B$ and with $A$ given by Eq.\eqref{warp-factor} the condition above is satisfied \textbf{(as can be seen in figures (\ref{fig.ia1}) and (\ref{fig.i1}))}. It is interesting to note that, in the domain wall case, this term is not present. That is why in 5D domain wall it is not possible to localize the vector field only by means of the gravitational interaction.

For further reference we may say that it is possible to obtain the zero mode localization for the vector field without imposing the gauge condition $A_{\theta} = 0$. Indeed, if one  consider a  $r$ dependence of $A_{\theta}$, say, $A_{\theta} = A_{\theta} (r)$, the system (\eqref{vector-eq1} - \eqref{vector-eq3}) will assume the form
\begin{eqnarray} \label{eq.vector-1}
\lefteqn{\left( \eta^{\mu\nu} \partial_{\mu}\partial_{\nu} + e^{B/2}\partial_{r} e^{-(A + \frac{B}{2})}\partial_{r} + \frac{e^{B - A}}{R_{0} ^{2}} \partial _{\theta} ^{2} \right) A_{\lambda} +} \nonumber\\
& & - e^{B/2} \left(\partial_{r} e^{-(A + \frac{B}{2})} \partial_{r} \right) \partial_{\lambda}A_{r} - \frac{e^{B - A}}{R_{0} ^{2}}  \partial_{\theta} \partial _{r} A_{\theta} = 0,
\end{eqnarray}
\begin{equation} \label{eq.vector-2}
\left(\eta^{\mu\nu}\partial_{\mu}\partial_{\nu} + \frac{e^{B - A}}{R_{0} ^{2}} \partial _{\theta} ^{2}\right) A_{r} - \frac{e^{B - A}}{R_{0} ^{2}}  \partial_{\theta} \partial _{r} A_{\theta} = 0,
\end{equation}
and
\begin{equation} \label{eq.vector-3}
\partial_{r}  (e^{-2 A + \frac{B}{2}}(\partial _{r} A_{\theta} - \partial _{\theta} A_{r}))  = 0.
\end{equation}

If the fields $A_{\lambda}$, and $A_{r}$ are decomposed as in Eq.\eqref{decomp-a-mi} and Eq.\eqref{decomp-a-r} it is possible to obtain the constant solution  $\rho_{m} = \rho_{0} = constant$ and $a_{r} = constant$ for the zero mode and s-wave. But in this case the function $A_{\theta}$ has to satisfy the equation
\begin{equation} \label{eq.a-teta}
A''_{\theta} (r) + \left(-2 A' + \frac{B'}{2}  \right) A'_{\theta}(r) = 0,
\end{equation}
whose general solution may be expressed as follows
\begin{equation} \label{sol-gen-a-teta}
A_{\theta} (r) = K_{1} \int ^{r} e^{2 A(r) - \frac{1}{2} B(r)} dr + K_{2},
\end{equation}
where $K_{1}, K_{2}$ are integration constants. Obviously this equation admits a more simple solution as $A_{\theta} = constant$. For this specific case the results obtained for the vector field localization would be the same found above for $A_{\theta} = 0$. However, as will be seen this constant solution will be relevant for the localization of the fermion field in the next section. It is worthwhile to mention here that other solutions for $A_{\theta}$ are possible, since they are finite for all $r$ \textit{and} $B(r)>4A(r)$. However the simplest solution to achieve fermion confinement is $A_{\theta}=constant$.

\section{Localization of the fermionic field\label{section4}}
\label{fermionic-local}

In this section, we will address the issue of fermionic field localization in a string-like defect scenario.
At first, we will following the procedure of the reference \cite{Oda2000}, where fermionic fields are localized on a string-like defect. However, as will be shown, it is not possible to localize the fermionic field in this scenario if one requires an exponentially decreasing warp factor, which is possible in the case of scalar and vector fields as we have seen.

On another standpoint there exist an approach developed by Liu and collaborators \cite{Liu2007}, where it is possible to localize the fermionic field
in the case of a decreasing warp factor. In this approach, one modifies the covariant derivative by adding a minimal coupling between the fields. On other hand, here we propose a mechanism in order to localize the fermion field using the minimal coupling in the bouncing brane keeping an decreasing warp factor in scenarios of codimension two brane.

To begin with, let us proceed as in the reference \cite{Oda2000}. We have that the fermionic field action can be written in a scenario
with six dimensions as
\begin{equation} \label{fermion-action}
S = \int {d^{6} x \sqrt{-g} \bar{\Psi} i \Gamma ^{M} D_{M} \Psi},
\end{equation}
and the equation of motion related with this action is
\begin{equation} \label{fermion-eq-motion}
\left( \Gamma ^{\mu} D_{\mu} + \Gamma ^{r} D_{r} + \Gamma ^{\theta} D_{\theta}\right) \Psi(x^{M}) = 0,
\end{equation}
where the matrices $\Gamma ^{M}$ represent the Dirac matrices in a curved spacetime. These matrices are related with the Dirac matrices in the flat spacetime as
\begin{equation} \label{gam-matr}
\Gamma ^{M} = h_{\bar{M}}^{M} \gamma ^{\bar{M}},
\end{equation}
where the  \textit{vielbein} $h_{\bar{M}}^{M}$ is given by the relation
\begin{equation} \label{vielbein}
g_{M N} = \eta _{\bar{M} \bar{N}}h_{M}^{\bar{M}} h_{N}^{\bar{N}}.
\end{equation}

The covariant derivative has the standard form
\begin{equation} \label{deri-covar}
D_{M} = \partial _{M} + \frac{1}{4} \Omega_{M} ^{\bar{M} \bar{N}} \gamma _{\bar{M}} \gamma _{\bar{N}},
\end{equation}
where the spin connection  $\Omega_{M} ^{\bar{M} \bar{N}}$ is defined by
\begin{eqnarray}
\label{spin-conx}
\lefteqn{\Omega_{M} ^{\bar{M} \bar{N}} = \frac{1}{2}h ^{N \bar{M}} \left(\partial _{M} h_{N} ^{\bar{N}} - \partial _{N} h_{M} ^{\bar{N}}\right) + }\nonumber\\
& & - \frac{1}{2}h ^{N \bar{N}} \left(\partial _{M} h_{N} ^{\bar{M}} - \partial _{N} h_{M} ^{\bar{M}}\right) - \frac{1}{2}h ^{P \bar{M} } h^{Q \bar{N}} h_{M} ^{\bar{R}} \left(\partial _{P} h_{Q \bar{R}} - \partial _{Q} h_{P \bar{R}}\right).
\end{eqnarray}

In order to write explicitly the equation \eqref{fermion-eq-motion}, we have to calculate the matrices $\Gamma ^{M}$
as well as the covariant derivative.  Using the metric \eqref{sl-metric} and the equation \eqref{gam-matr}, we have that
the relation between the Dirac matrices in a curved space-time and the Dirac matrices in the flat space-times is given by
\begin{equation} \label{gam-matr2}
\Gamma ^{\mu} = P^{-\frac{1}{2}} \gamma ^{\bar{\mu}}; \hspace{5 pt}  \Gamma ^{r} =  \gamma ^{\bar{r}}; \hspace{5 pt} \Gamma ^{\theta} = Q^{-\frac{1}{2}} \gamma ^{\bar{\theta}}.
\end{equation}

The nonvanishing components of the spin connection \eqref{spin-conx} are
\begin{equation} \label{spin-conx-nonvan}
\Omega_{\mu } ^{\bar{r} \bar{\mu }} = - \frac{1}{2} P^{-\frac{1}{2}} P' \delta _{\mu} ^{\bar{\mu }}; \hspace{5 pt} \Omega_{\theta} ^{\bar{r} \bar{\theta}} = - \frac{1}{2} Q^{-\frac{1}{2}} Q' \delta _{\theta} ^{\bar{\theta }}.
\end{equation}
Moreover, one can explicitly write the covariant derivative components \eqref{deri-covar} as
\begin{equation} \label{derv-cov-corda}
D_{\mu} \Psi = \left( \partial _{\mu} - \frac{1}{4} \frac{P'}{P}  \Gamma _{r} \Gamma _{\mu} \right) \Psi; \hspace{5 pt} D_{\theta} \Psi = \left( \partial _{\theta} - \frac{1}{4} \frac{Q'}{Q}  \Gamma _{r} \Gamma _{\theta} \right) \Psi; \hspace{5 pt} D_{r} \Psi =  \partial _{r} \Psi.
\end{equation}

In order to write  the equations of motion to these fields, we have to set the way how the Dirac matrices act on the \textit{spinor} $\Psi$.
This approach was presented by the references \cite{Oda2000,Liu2007,Davi}, which we follow closely. First, let us assume that
the spinor can be written in two parts, the \textit{right} part $\Psi _{R}$ and the \textit{left} part $\Psi _{L}$ as
\begin{equation} \label{spinor-decomp-corda}
\Psi(x^{M}) = \sum_l(\Psi _{R} \alpha _{R} + \Psi _{L} \alpha _{L}) e^{i l \theta}\;.
\end{equation}

The $\Gamma$ matrices act on these spinor as
\begin{equation} \label{spinor-matrice-corda}
\Gamma ^{\mu} \partial _{\mu} \Psi _{R} (x^{\mu}) = m \Psi _{L} (x^{\mu}); \hspace{12 pt} \Gamma ^{\mu}  \partial _{\mu} \Psi _{L} (x^{\mu})
= m \Psi _{R} (x^{\mu})\;.
\end{equation}
Or yet in terms of the $\gamma$ matrices in the flat spacetime, as
\begin{equation}\label{psimass}
\gamma^{\mu}\partial_{\mu}\Psi_R(x^{\mu})= P^{-\frac{1}{2}}m\Psi_L(x^{\mu})\;\; ; \;\;
\gamma^{\mu}\partial_{\mu}\Psi_L(x^{\mu})= P^{-\frac{1}{2}}m\Psi_R(x^{\mu}).
\end{equation}

Naturally, for $m = 0$ one has
\begin{equation}
\gamma ^{\mu} \partial _{\mu} \Psi _{R} (x^{\mu}) = \gamma ^{\mu} \partial _{\mu} \Psi _{L} (x^{\mu}) = 0\;.
\end{equation}

These equations can still be put in the form
\begin{equation} \label{spinor-matrice-corda2}
\gamma ^{r} \Psi _{R} (x^{\mu}) = + \Psi _{R} (x^{\mu}); \hspace{12 pt} \gamma ^{r}   \Psi _{L} (x^{\mu}) = - \Psi _{L} (x^{\mu}),
\end{equation}
\begin{equation} \label{spinor-matrice-corda3}
\gamma ^{\theta} \Psi _{R} (x^{\mu}) = i \Psi _{R} (x^{\mu}); \hspace{12 pt} \gamma ^{\theta}   \Psi _{L} (x^{\mu}) = i \Psi _{L} (x^{\mu}).
\end{equation}

We require here that $\psi (x^{\mu})$ must satisfies the Dirac equation on the brane, namely $\gamma ^{\mu} \psi _{\mu} = 0$. Thus, taking into account the s-wave case, and the equations \eqref{gam-matr2}, \eqref{spin-conx-nonvan}, \eqref{spinor-decomp-corda}, \eqref{spinor-matrice-corda},
\eqref{spinor-matrice-corda2} and \eqref{spinor-matrice-corda3}, we will have that the equations of motion \eqref{fermion-eq-motion} can be written as
\begin{equation} \label{eq-fermion-corda}
\left( \partial _{r} + \frac{P'}{P} + \frac{1}{4}\frac{Q'}{Q} \right) \alpha (r) = 0.
\end{equation}

The general solution to the equation above is given by
\begin{equation} \label{fermion-sol-eq-corda}
\alpha (r) = c_{2} P^{-1} Q^{- \frac{1}{4}},
\end{equation}
where $c_{2}$ is a constant of integration.

It is still necessary to verify if the solution is normalizable. In order to do this, we need analyze the action \eqref{fermion-action}
with the spinor $\psi$ replaced by this solution and verify if the resultant integral in the variable $r$ is finite.
From this, the interesting integral for the case analyzed here assumes the following form
\begin{equation} \label{fermion-action-corda}
I_{\frac{1}{2}}  \propto \int _{0} ^{\infty} dr P^{\frac{3}{2}} Q^{\frac{1}{2}} \alpha (r) ^{2}.
\end{equation}
At last, let us replace the expression \eqref{fermion-sol-eq-corda} for $\alpha$ in the equation \eqref{fermion-action-corda} in a way that, for the case
where $A = B$, we have
\begin{equation} \label{fermion-action-corda1}
I_{\frac{1}{2}}  \propto  \int _{0} ^{\infty} dr P^{\frac{3}{2}} P^{- 2}  \propto \int _{0} ^{\infty} dr e^{\frac{1}{2} A(r)}.
\end{equation}
We note that no localization of the fermionic fields is possible in the case of a smooth warp factor given by
$A(r) = \beta \ln \cosh  ^{2}(ar) + \frac{\beta}{2} \tanh^{2}(ar)$. Even the possibility of assume $\beta < 0$  is not possible since
$\beta = \frac{1}{3} \kappa_{6} ^{2} \nu ^{2}$.  In consequence, in our case it is not possible to appeal to a growing warp factor.

Therefore, the only alternative that remains is to use the treatment introduced in the reference \cite{Liu2007}
which consists in modify the covariant derivative \eqref{deri-covar} by adding a term of minimal coupling.
If one does this, the new covariant derivative reads
\begin{equation} \label{deri-covar-acopla}
D_{M} = \partial _{M} + \frac{1}{4} \Omega_{M} ^{\bar{M} \bar{N}} \gamma _{\bar{M}} \gamma _{\bar{N}} - i e A_{M},
\end{equation}
where $A_{M}$ is a gauge field, $e$ the electrical charge, and $i$ is the imaginary unit.
Such modification does not change the relation between the gamma matrices \eqref{gam-matr2}, neither the nonvanishing components
of the spin connection \eqref{spin-conx-nonvan}. On the other hand, the covariant derivatives components in this case, will be changed to the following forms
\begin{eqnarray} \label{derv-cov-corda2}
D_{\mu} \Psi &=& \left[ \partial _{\mu} -  (P'/8P)  \Gamma _{r} \Gamma _{\mu} - i e A{\mu} \right] \Psi,  \nonumber \\
\\ \nonumber
D_{r} \Psi &=&  \left( \partial _{r} - i e A{r}\right) \Psi,
\\
\\ \nonumber
D_{\theta} \Psi &=& \left[ \partial _{\theta} -  (Q'/8Q)  \Gamma _{r} \Gamma _{\theta} - i e A_{\theta} \right] \Psi.
\end{eqnarray}

Using the conditions \eqref{spinor-decomp-corda}, \eqref{spinor-matrice-corda}, \eqref{spinor-matrice-corda2} and \eqref{spinor-matrice-corda3},
taking into account only the zero mode of the field and for the s-wave, the equation of motion for the \textit{right mode} reads
\begin{equation} \label{eq-fermion-corda2}
\left( \partial _{r} + \frac{P'}{P} + \frac{1}{4}\frac{Q'}{Q} -i e A_{r} (r) + e Q^{-\frac{1}{2}} A_{\theta} (r) \right) \alpha (r) = 0,
\end{equation}
where we have assumed that the Dirac equation $\bar{\gamma}^{\mu} \partial_{\mu} \Psi(x^{\mu}) = 0$ is valid on the brane,
and the field $A_{M}$ can be decomposed in its  components $A_{\mu} (x^{\mu})$, $A_{r} (r)$, and $A_{\theta} (r)$.
The solution of the equation  \eqref{eq-fermion-corda2} is given by
\begin{equation} \label{fermion-sol-eq-corda2}
\alpha (r) = c_{3} P^{-1} Q^{- \frac{1}{4}} \exp \left(\int ^{r}( i e A_{r} - e Q^{-\frac{1}{2}} A_{\theta}) dr \right).
\end{equation}

Inserting this solution in the action \eqref{fermion-action}, the integral in the variable $r$ reads
\begin{eqnarray} \label{fermion-action-corda3}
I_{\frac{1}{2}}  &\propto&  \int _{0} ^{\infty} \left( dr P^{-\frac{1}{2}} \exp \left( - 2 e \int ^{r}  Q^{-\frac{1}{2}} A_{\theta} \right) \right)  = \nonumber \\
&=&\int _{0} ^{\infty} dr \exp \left(  \frac{1}{2} A(r)  - 2 e R_{0} ^{-1} \int ^{r} e^{\frac{1}{2} B (r)} A_{\theta} \right).
\end{eqnarray}

The form of field $A_{\theta}$ in the above integral is essential to ensure the zero mode localization, i.e,
to ensure the finiteness of the integral. In this case it is sufficient that  $A_{\theta} = a_{\theta} = constant$.
The integral  (\eqref{fermion-action-corda3}) is naturally convergent for $r \rightarrow 0$, so we have to study it in the other limit,
$r \rightarrow \infty$. It is easy to see that in this limit the function $A(r)$ is linear on $r$ and, in this case, it is possible
to write $A(r)=B(r) \approx \beta r$. In this situation we can easily see that (\eqref{fermion-action-corda3}) converges. Indeed, it will assumes the simple form
\begin{equation} \label{fermion-action-corda-4}
I_{\frac{1}{2}}  \propto   \int _{0} ^{\infty} dr \exp \left(  \frac{1}{2} \beta r  - \frac{4a_0}{R_0\beta}  e^{\frac{1}{2} \beta r} \right).
\end{equation}

\begin{figure}[htb] %
       \begin{minipage}[t]{0.48 \linewidth}
                  \fbox{\includegraphics[width=\linewidth]{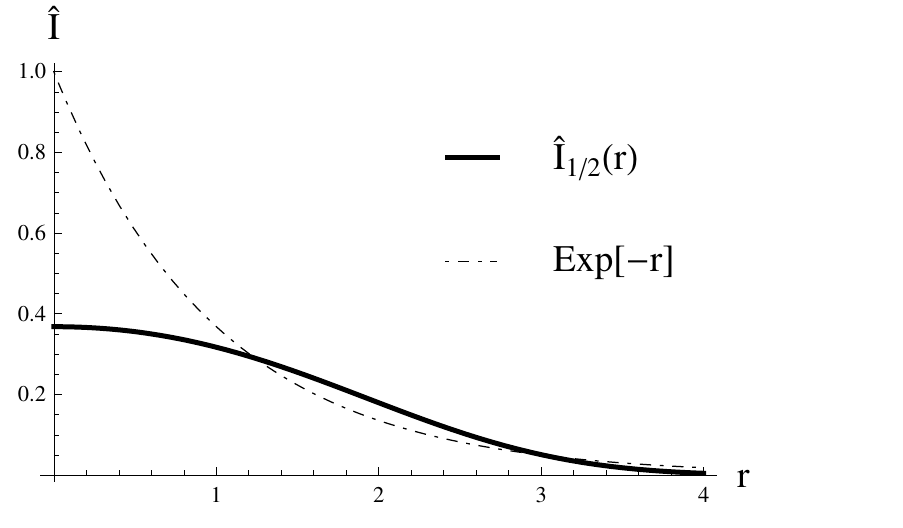}}\\
           \caption{Sketch of the integrand in eq.\eqref{fermion-action-corda-4}. We can verify that this is a case of a smooth linear decreasing exponential. Here $\beta=a=1$ and $a_0=2$.}
           \label{fig.ia12}
       \end{minipage}\hfill
       \begin{minipage}[t]{0.48 \linewidth}
        \fbox{\includegraphics[width=\linewidth]{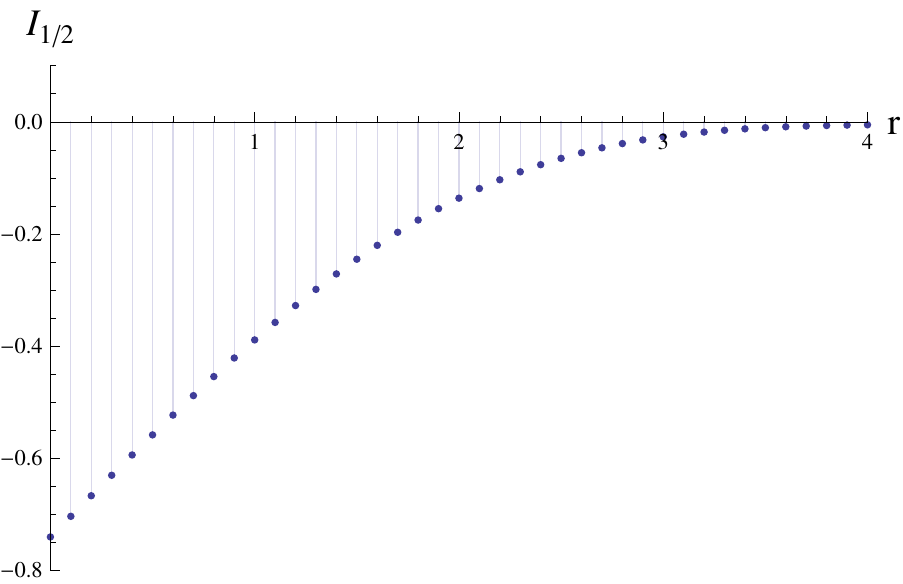}}\\
        \caption{Proof of convergence of $I_{\frac{1}{2}}$ (Eq.\eqref{fermion-action-corda-4}). The function $I_{\frac{1}{2}}$ has a form similar to a linear decreasing exponential. After integration $I_{\frac{1}{2}}$ worth approximately $0.736$ for $\beta=a=1$ and $a_0=2$.}
         \label{fig.i12}
       \end{minipage}
\end{figure}

Now, setting $a_0\equiv\frac{R_0\beta}{8}$ and since the exponential function always assumes values greater than values from a linear function, one sees that the integral above is convergent (as can be seen in figures (\ref{fig.ia12}) and (\ref{fig.i12})). Moreover, for left handed fermions we must set $a_0\to-a_0$ in order to converge that integral.

%%%%%%%%%%%%%%%%%%%%%%%%%Figures%%%%%%%%%%%%%%%%%%%%%%%%%%%%%%%%%%%%%%%%%%%%%
%           \fbox{\includegraphics[width=\linewidth]{Ia12e.eps}}\\
%         \caption{Integrando de \eqref{fermion-action-corda1}. Notamos ser divergente quando $r$ tende ao infinito.}
%         \label{fig.ia12e}

\section{Fermionic massive modes\label{section4b}}

In this section we treat the fermionic massive modes using a qualitative analysis. In order to do this, we put the equations of
motion of the massive spinor in the form of the Schr\"{o}dinger equation.

First, we consider the massive counterpart of the eq.\eqref{eq-fermion-corda2}. Indeed, with help of eq.\eqref{psimass} we arrive at
\begin{eqnarray}
\label{fermion-mass}
\left[ \partial _{r} + H_{R,L}(r)\right]\alpha_{R,L}(r)=\pm m P^{-\frac{1}{2}}\alpha_{L,R} (r),
\end{eqnarray}
 where $H_{R,L}(r)=\frac{P'}{P} + \frac{1}{4}\frac{Q'}{Q} -i e A_{r} (r) \pm e Q^{-\frac{1}{2}} A_{\theta} (r) $. Note that the expression \eqref{fermion-mass} is composed by two equations with coupled quiralities. Applying the following change in the independent variable $\frac{dz}{dr}=P^{-\frac{1}{2}}(r)$ and decoupling the left mode and the right mode we have
\begin{eqnarray}
\label{fermion-mass2}
\left[ \partial _{z} + H_{R,L}(z)\right]\left[ \partial _{z} + H_{L,R}(z)\right]\alpha_{R,L}(z)=-m^2\alpha _{R,L}(z).
\end{eqnarray}

Through a new change of variable $\tilde{\alpha}_{R,L}(z)=\exp\left[-\int_z H_{R,L}(z)dz\right]\alpha_{R,L}(z)$, we can put eq.\eqref{fermion-mass2} in the form of Schr\"{o}dinger equation, namely
\begin{eqnarray}
\label{fermion-mass3}
\left(-\partial_{z}^{2}+ \left[-\partial_z\left(e \frac{A_{\theta}(z)\sqrt{P(z)}}{R_0}\right)\mp\left(e  \frac{A_{\theta}(z)\sqrt{P(z)}}{R_0}\right)^2\right] \right)\tilde{\alpha}_{R,L}(z)=m^2\tilde{\alpha}_{R,L}(z).
\end{eqnarray}
Therefore our potential can be written in term of $r$ as
\begin{eqnarray}\label{vbounce}
V_{R,L}(r)=\left[-\frac{e}{R_0}\sqrt{P(r)}\partial_r\left( A_{\theta}(r)\sqrt{P(r)}\right)\mp\left(e  \frac{A_{\theta}(r)\sqrt{P(r)}}{R_0}\right)^2\right] .
\end{eqnarray}

\begin{figure}[htb]
\centering
           \fbox{\includegraphics[scale=.7]{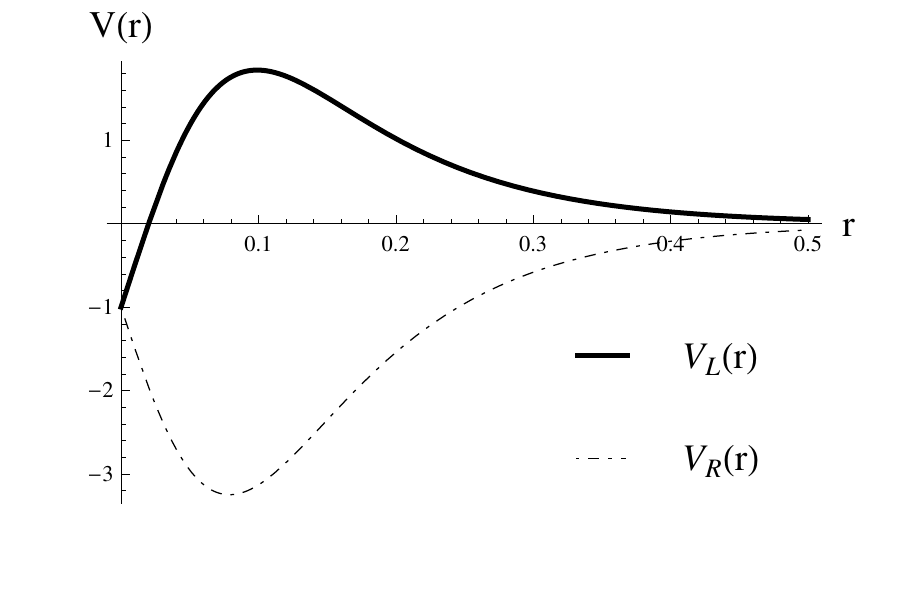}}\\
         \caption{Sketch of the left mode potential (thick line) and sketch of the right mode potential (dot-dashed line) of the eq.\eqref{vbounce}. Here $\beta=1$ and $a=10$.}
         \label{Fig.vbounce}
 \end{figure}

Now we study the choice $A_{\theta}(r)= constant$ which it was convenient for confinement of the zero modes of the vectorial and spinorial fields. For this choice, we present a sketch of the potentials in Fig.(\ref{Fig.vbounce}).

We note from the fig.(\ref{Fig.vbounce}) that both potentials are gap free when $r\to\infty$. Also we note that only the left mode potential assumes a volcano type form. This feature indicates that only left handed fermions are confined in our brane. As a matter of fact, the potential for right handed fermions is always attractive. Also it reaches their asymptotic value on the infinity under the axis, which would not guarantee the right mode localization \cite{Zhao:2009ja}.

The complete analysis of the massive modes must to include the calculation of the resonant modes. However this calculation requires the numerical solution of equation \eqref{fermion-mass3}. We defer this numerical analysis for a future work.

\section{Remarks and conclusions} \label{section5}

This work adds results in studies about thick brane in codimension two spaces. Here we have implemented a mechanism in order to localize vector and fermion fields in the scenario introduced in Ref. \cite{Sousa:2012jw}, where a thick brane is generated from a scalar field on a string-like defect. We have found that both, vector and fermion fields, can be localized
in this scenario only with the gravitational interaction, which confirms the applicability and robustness of this model.

It is worthwhile to mention that in order to localize fermion field we show that, for the first time treating a bouncing brane, a previous localization of the gauge field is required. Indeed, we must have a component of the gauge field in the direction of the angular extra coordinate to obtain a convergent result for the integral of localization. In the literature, this component is usually turns to be zero as a gauge choice \cite{Liu2007}.

In the case under analysis here it was not necessary to appeal to a growing warp factor in order to localize fermion fields, since no localization of these fields is possible in the case of the smooth warp factor \eqref{warp-factor}. Therefore, the only alternative in order to localize fermions in this context it is to modify the covariant derivative \eqref{deri-covar} by adding a term of minimal coupling \cite{Liu2007}, which was never used for the bouncing brane. In this case, the fermionic field can be localized on the brane.

Finally, we perform a qualitative analysis of the fermionic massive modes. We conclude that only left handed fermions could be localized in the brane. A numerical analysis of the massive modes is deferred for a future work.

\section*{Acknowledgments}

The authors thank the Funda\c{c}\~{a}o Cearense de apoio ao Desenvolvimento Cient\'{\i}fico e Tecnol\'{o}gico (FUNCAP), the Coordena\c{c}\~{a}o de Aperfei\c{c}oamento de Pessoal de N\' ivel Superior (CAPES), and the Conselho Nacional de Desenvolvimento Cient\' ifico e Tecnol\' ogico (CNPq) for financial support.

\end{document}